\documentclass[fleqn,usenatbib]{mnras}

\usepackage{newtxtext,newtxmath}

\usepackage[T1]{fontenc}

\DeclareRobustCommand{\VAN}[3]{#2}
\let\VANthebibliography\thebibliography
\def\thebibliography{\DeclareRobustCommand{\VAN}[3]{##3}\VANthebibliography}

\usepackage{graphicx}	
\usepackage{amsmath}	

\title[Capture of halo material by subhaloes]{The capture of halo material by orbiting subhaloes}

\author[Hang Yang et al.]{
Hang Yang,$^{1,2,3,4}$\thanks{E-mail: hyang@nao.cas.cn}
Simon D.M. White,$^{2}$\thanks{E-mail: swhite@mpa-garching.mpg.de}
and Liang Gao$^{1,3,4}$
\\
$^{1}$ Institute for Frontiers in Astronomy and Astrophysics, Beijing Normal University, Beijing 102206, China \\
$^{2}$ Max Planck Institute for Astrophysics, Karl-Schwarzschild-Strasse 1, 85740 Garching, Germany \\
$^{3}$ School of Physics and Astronomy, Beijing Normal University, Beijing 100875, China \\
$^{4}$ Institute for Astrophysics, School of Physics, Zhengzhou University, Zhengzhou 450001, China \\
}

\date{Accepted XXX. Received YYY; in original form ZZZ}

\pubyear{\the\year{}}

\begin{document}
\label{firstpage}
\pagerange{\pageref{firstpage}--\pageref{lastpage}}
\maketitle

\begin{abstract}
When a dark matter halo falls into a more massive object and becomes a subhalo, it typically loses much of its mass through tidal stripping. The reverse process is also possible in principle. The  subhalo may gravitationally capture material from its host. If sufficiently efficient, this process could make an initially starless subhalo visible. We use high-resolution N-body simulations to estimate the efficiency of capture. We find that after an extended period orbiting within its host, at most a few times $10^{-4}$ of a subhalo's remaining mass has been acquired since infall.
This captured material is less concentrated to subhalo centre than material retained from before infall. It is also very much less abundant than host material that is instantaneously passing through the subhalo on almost unperturbed orbits.
Captured stars are not sufficiently spatially concentrated to be distinguished from the dominant  background of "field" stars, and their concentration in velocity space is no greater than that of typical stellar streams in the halo. Unfortunately, stellar capture is not efficient enough to allow initially starless low-mass subhaloes to be detected.


\end{abstract}




\section{Introduction}
In the standard $\Lambda$-cold dark matter ($\Lambda$CDM) paradigm,  cosmic structure grows hierarchically. Small dark matter haloes collapse and come to equilibrium first, thereafter growing in mass by smooth accretion and by merging with other haloes. During this process, accreted haloes are not fully disrupted within their new and more massive hosts; rather their inner regions survive as self-bound subhaloes orbiting within the larger system \citep[e.g.][]{1998ApJ...499L...5M, 2004MNRAS.355..819G, 2008MNRAS.391.1685S}.  

The mass of the smallest dark matter halo depends on the nature of the dark matter but, for cold dark matter, is assumed to be much smaller than galactic scale. For example, for weakly interacting massive particles (WIMPs) it is set by the damping scale due to early free-streaming, about $10^{-6} \rm M_{\odot}$ for a rest mass of 100 GeV/c$^2$ \citep{2005PhR...405..279B}, Both analytical models \citep[e.g.][]{1974ApJ...187..425P, 1991ApJ...379..440B} and numerical simulations \citep[e.g.][]{1999MNRAS.308..119S, 2001MNRAS.321..372J, 2020Natur.585...39W, 2024MNRAS.528.7300Z} have shown that at low mass the present-day abundance of dark matter haloes in a $\Lambda$CDM cosmology increases almost as ${\rm d}N/{\rm d}M \propto M^{-2}$. This is much steeper than the observed abundance of galaxies as a function of stellar mass, so most low-mass haloes (and subhaloes) must be without stars \citep[e.g.][]{1999ApJ...522...82K, 2010MNRAS.404.1111G, 2015MNRAS.448.2941S}. 

In hierarchical models like $\Lambda$CDM, galaxies form by the cooling and condensation of gas at the centres of the growing population of dark matter haloes  \citep{1978MNRAS.183..341W}. It is generally believed that only haloes capable of efficient atomic cooling can sustain star formation and so build substantial stellar systems. This requires a virial temperature larger than about $10^{4}$ K. In addition, after reionization ($z<6$) UV photons prevent gas in low-mass haloes from condensing and forming stars \citep[e.g.][]{1992MNRAS.256P..43E, 2008MNRAS.390..920O, 2023MNRAS.525.5932B, 2024arXiv241016176Z}. Considering gas cooling and photoheating together, there is a lower limit of order $10^9M_\odot$  on the mass of present-day haloes that can host a sigificant galaxy \citep[e.g.][]{2020MNRAS.498.4887B,2022ApJ...941..120G}. Below we will use the term "starless" (sub)haloes to refer to haloes below this star formation limit and to the subhaloes formed when they accrete onto a larger system.

Recently, \cite{2024MNRAS.533.3263P} argued that starless subhaloes may become visible by  capturing field stars from the stellar halo of their host. If this mechanism is significant, it could make low-mass subhalo populations detectable, thus confirming their existence and giving important insight into the nature of dark matter. As a result, it is of considerable interest to explore this process directly in order to estimate the capture efficiency. Since the capture process for field stars is identical to that for individual dark matter particles on the same orbit, we can estimate the required efficiency by evaluating the frequency with which simulation particles are captured by orbiting suhaloes in dark-matter-only simulations..

To formalise this argument we note that in collisionless stellar dynamics, the probability that a test particle is captured by an orbiting substructure depends only on its orbit and is independent of whether it represents a dark matter particle, a star or a simulation particle. Orbits within an evolving collisionless system can be characterised by their phase-space coordinate $\bf{\tau_0} = (\bf{x_0}, \bf{v_0})$ at any chosen fiducial time
$t_0$. The mass of stars captured by a population of orbiting substructures can then be written as $M_*=\int d^6\tau_0~ p(\tau_0)f_*(\tau_0)$ where $f_*$ is the (conserved) phase-space density of stars, and $p$ takes the value 1 if an orbit is captured and is zero otherwise. We then have 
\begin{align}
M_* &= \int d^6\tau_0~(f_*/f_{\rm dm})p(\tau_0)f_{\rm dm}(\tau_0)\nonumber\\ & = \langle m_*/m_{\rm dm}\rangle\int d^6\tau_0~p(\tau_0)f_{\rm dm}(\tau_0)\nonumber\\ & = \langle m_*/m_{\rm dm}\rangle M_{\rm dm}
\label{eq:MLarg}
\end{align}
where $M_{\rm dm}$ is the captured dark matter mass and $\langle m_*/m_{\rm dm}\rangle$ is the star to dark matter mass ratio averaged over all captured orbits weighted by the dark matter mass on each orbit. Since subhaloes have shallow potential wells, capture requires test particle and subhalo to have similar position and velocity (hence similar orbits) at the time of capture.  Hence we will estimate  $\langle m_*/m_{\rm dm}\rangle$ observationally as the stellar to dark matter mass ratio in the region where subhaloes currently orbit (hence effectively taking $t_0$ to be the present).  In Appendix B we give a simple but moderately realistic toy model which illustrates this argument

The paper is organised as follows. In Section 2 we describe the numerical simulations and our analysis methods. The main results are presented in Section 3, and we summarise and discuss them in Section 4. Appendix A demonstrates that, appropriately expressed, particle capture by subhaloes is independent both of halo mass scale (over a factor of 1000 in halo mass) and of simulation resolution (over a factor of 8 in mass resolution). Appendix B presents a toy model supporting the argument surrounding equ.~(\ref{eq:MLarg}) and discusses further aspects of this argument.


\section{Methods}
\subsection{The Phoenix Simulations}

In this paper, we use the dark-matter-only, zoom-in simulations of the Phoenix suite \citep{2012MNRAS.425.2169G} to explore whether field particles in a halo can be captured by subhaloes. The collisionless, gravitational dynamics of dark matter particles and of field stars are identical, so we do not need to identify ``stars'' explicitly in the simulations and can concentrate instead on the capture of dark matter particles.
The Phoenix suite resimulated the evolution of nine galaxy cluster regions selected from the Millennium Simulation \citep{2005Natur.435..629S}. These resimulations were run using GADGET-3, an updated version of the publicly available GADGET-2 code \citep{2005Natur.435..629S},  adopting WMAP1 cosmological parameters: $\Omega_m=0.25$, $\Omega_b=0.048$, $\Omega_{\Lambda}=0.75$, $h=0.73$, $n_s=1$ and $\sigma_8=0.8$ \citep{2003ApJS..148..175S}. The Friends-of-Friends \citep[FOF,][]{1985ApJ...292..371D} and SUBFIND \citep[]{2001MNRAS.328..726S} algorithms were applied to identify halo and subhalo structures, respectively. The Phoenix simulations were run at several mass resolution levels. Here, we use only the level-2 results, the highest resolution available for all nine haloes. These have mass resolution $\sim 10^6 \rm M_{\odot}$ which, since the halo virial masses range  from $7 \times 10^{14} \rm M_{\odot}$ to $3 \times 10^{15}\rm M_{\odot}$, corresponds to approximately $10^8$ particles within the virial radius $R_{200}$ of each $z=0$ halo.\footnote{We define the virial radius $R_{200}$ as the radius within which the mean density is 200 times the critical density of the Universe.} 

\citet{2012MNRAS.425.2169G} showed explicitly and in detail that the evolution of the Phoenix cluster mass haloes closely resembles a scaled up version of that in the Aquarius suite of Milky-Way mass haloes \citep{2008MNRAS.391.1685S}. Thus we believe that the results we find below should be applicable to halo star capture in the Milky Way's halo once all masses are scaled down by a factor of 1000 and all distances and velocities by a factor of 10. We check this point explicitly in Appendix A by repeating some of our analysis for the two highest resolution simulations of one of the Aquarius haloes.

\subsection{Sample Selection and Analysis Method}

Captured particles will be in, or close to a subhalo and will move with it at late times, but they will not be bound to or otherwise associated with it at the time of its accretion. We use the following procedures to identify all such newly associated subhalo particles in each of our Phoenix simulations

1) We track each surviving $z=0$ subhalo with at least 100 bound particles back to the latest simulation snapshot in which its progenitor lies outside $R_{200}$ of the host halo. We then define the infall redshift of the subhalo $z_{\rm inf}$ as that of the next stored snapshot, and the subhalo's bound particle number at infall $n_{\rm inf}$ to be the value found at $z_{\rm inf}$. 

2) We find the latest snapshot, $z_{\rm max}>z_{\rm inf}$, for which the subhalo's progenitor is the main subhalo of its FoF group, and we measure its $R_{200}$ at this time. We then go to the snapshot immediately before $z_{\rm inf}$ and record the IDs of all particles that are within $r_{\rm limit}$ of halo centre in this snapshot, taking $r_{\rm limit} = 2R_{200}(z_{\rm max})$. For the $i$-th $z=0$ subhalo we denote this set of particles as $A_i$. Our results below are insensitive to the precise definition of $r_{\rm limit}$. 

3) At $z\approx 0.1$, we record the particle IDs of all particles located within twice the tidal radius, $2r_t$ of the centre of each subhalo. We denote this set of particles as $B_i$. We approximate the tidal radius of a subhalo of SUBFIND mass $M_{\rm sub}$ using the circular orbit formula,
\begin{equation}
\label{label:r_t}
r_t = r [\frac{M_{\rm sub}}{(2 - \partial{\ln M}/\partial{\ln r})M(<r)}]^{1/3},
\end{equation}
where $r$ is the distance from main halo centre and $M(<r)$ is the mass enclosed within $r$ \citep[e.g.][]{1998MNRAS.299..728T, 2008MNRAS.391.1685S}.

4) We apply the same procedure as in step (3) at $z=0$ and define the corresponding set of particles as $C_i$.

5) For the $i$-th subhalo, the particles in the set $B_i \setminus A_i$ (those in $B_i$ but not in $A_i$) are defined as newly associated particles at $z=0.1$. Note that we do not require these particles to be bound to the subhalo so they are a superset of captured particles and include particles which are just ``passing through''. Particles that belong to $A_i\cap B_{i}$ (those in both $A_i$ and $B_i$) are referred to as retained particles.

6) For each subhalo, we also test whether the sets $B_{i} \setminus A_{i}$ and $C_i$ share any elements. Particles that belong to both sets are defined as still associated and can be considered as truly captured particles.

In the following analysis, we drop from consideration a small number of subhaloes which experience a merger after their most recent infall. Such mergers could cause starless subhaloes to acquire stars in the so-called ``ghost galaxy'' scenario \citep{2023ApJ...958..166W}, but here we focus only on the smooth capture mechanism discussed by \citet{2024MNRAS.533.3263P}. 

\begin{figure*}
	\includegraphics[width=\textwidth]{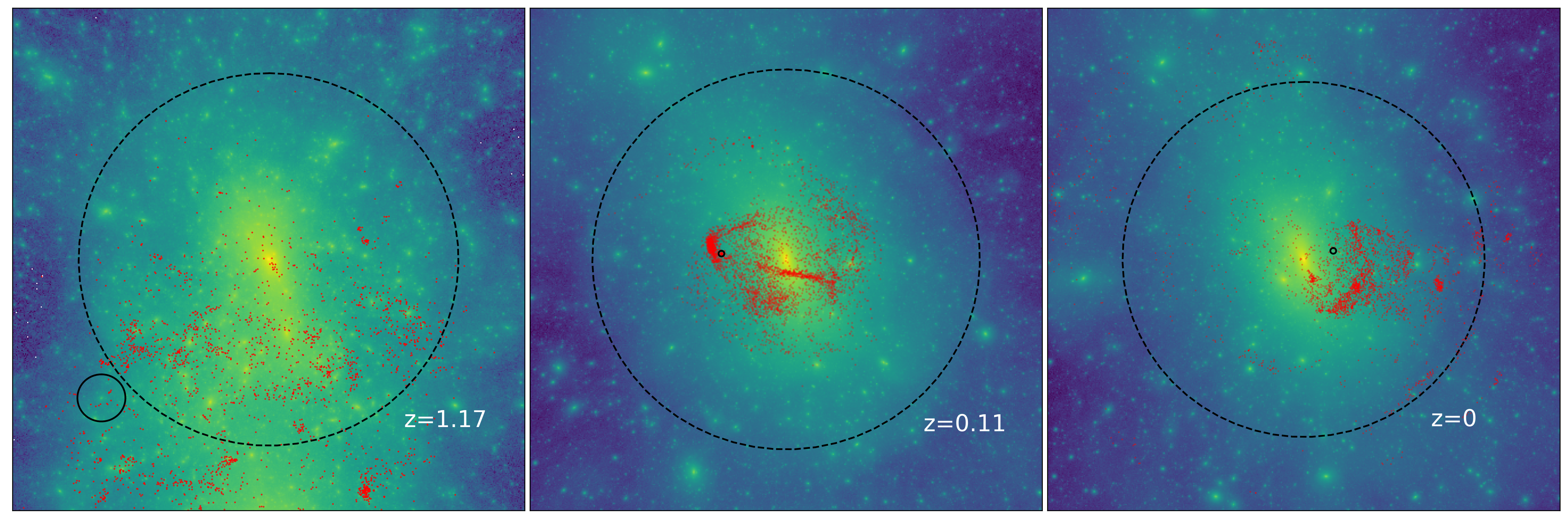}
	\centering
    \caption{An illustration of our analysis procedure for a randomly selected but relatively massive subhalo with $z_{\rm inf}=1.08$. From left to right, the panels show the projected distribution (in comoving coordinates) of mass and of various specific particle sets at $z=1.17, 0.11$ and 0.0 when sets A, B and C are selected, respectively. The black dashed circles show the virial radius of the main halo in each panel. The solid black circle in the left panel shows $r_{\rm limit}$ for the infalling halo. The red dots in this panel show the 2127 particles that will be considered newly associated to the subhalo at $z=0.11$. The solid black circles in the middle and right panels indicate  $2r_{t}$ for the subhalo. Red dots in the middle panel show particles that were within $r_{\rm limit}$ in the left panel. For clarity, we display only a random 10\% of these 41638 particles; 5\% of them are within $2r_{t}$ at $z=0.11$.The red dots in the right panel show the $z=0$ positions of particles that were newly associated at $z=0.11$ (the same particles plotted in red in the left panel). Each panel is 4cMpc/h on a side.}
    \label{fig:sketch}
\end{figure*}

\section{Results} \label{sec:res}

\subsection{An illustrative example}

Fig.~\ref{fig:sketch} presents an illustrative example, a randomly selected $z=0$ subhalo which was accreted at $z_{\rm inf} = 1.08$.  In the left panel, which shows the last snapshot in which this object was outside $R_{200}$ of its host halo, the black solid circle represents $r_{\rm limit}$ for the subhalo's progenitor. The 41638 particles within the corresponding sphere make up set $A$ as defined in Section 2.2. In the middle panel, the black solid circle represents $2r_{t}$ for the subhalo at $z=0.11$, so the 2127 particles within this sphere define set $B$. Similarly, the black solid circle in the right panel represents $2r_{t}$ for the subhalo at $z=0$, so the 2348 particles within this sphere define set $C$. The red dots in the left panel indicate the 1308 particles that will be considered newly associated at $z=0.1$, and hence belong to set $B \setminus A$. The red dots in the middle panel show 10\% of the particles in set $A$ at $z=0.1$, while the red dots in the right panel show the 1308 particles in $B\setminus A$ at $z=0$. None of these is still inside $2r_t$ at this time, so $(B\setminus A)\cap C$ is empty.

As the middle panel shows, most particles associated with the infalling halo (95\% of set $A$) are no longer close to the subhalo at $z=0.11$, but are strewn around the main halo as a result of tidal disruption and mixing processes. There are indeed 1308 particles that are within $2r_t$ of the subhalo at $z=0.11$ but were not close to it at infall (the red dots in the left panel). However, as shown by the red dots in the right panel, none of these newly associated particles is still within $2r_t$ at $z=0$, just $\sim 1$ Gyr later. Thus none of the newly associated particles was permanently captured by the subhalo.

\begin{figure*}
	\includegraphics[width=0.8\textwidth]{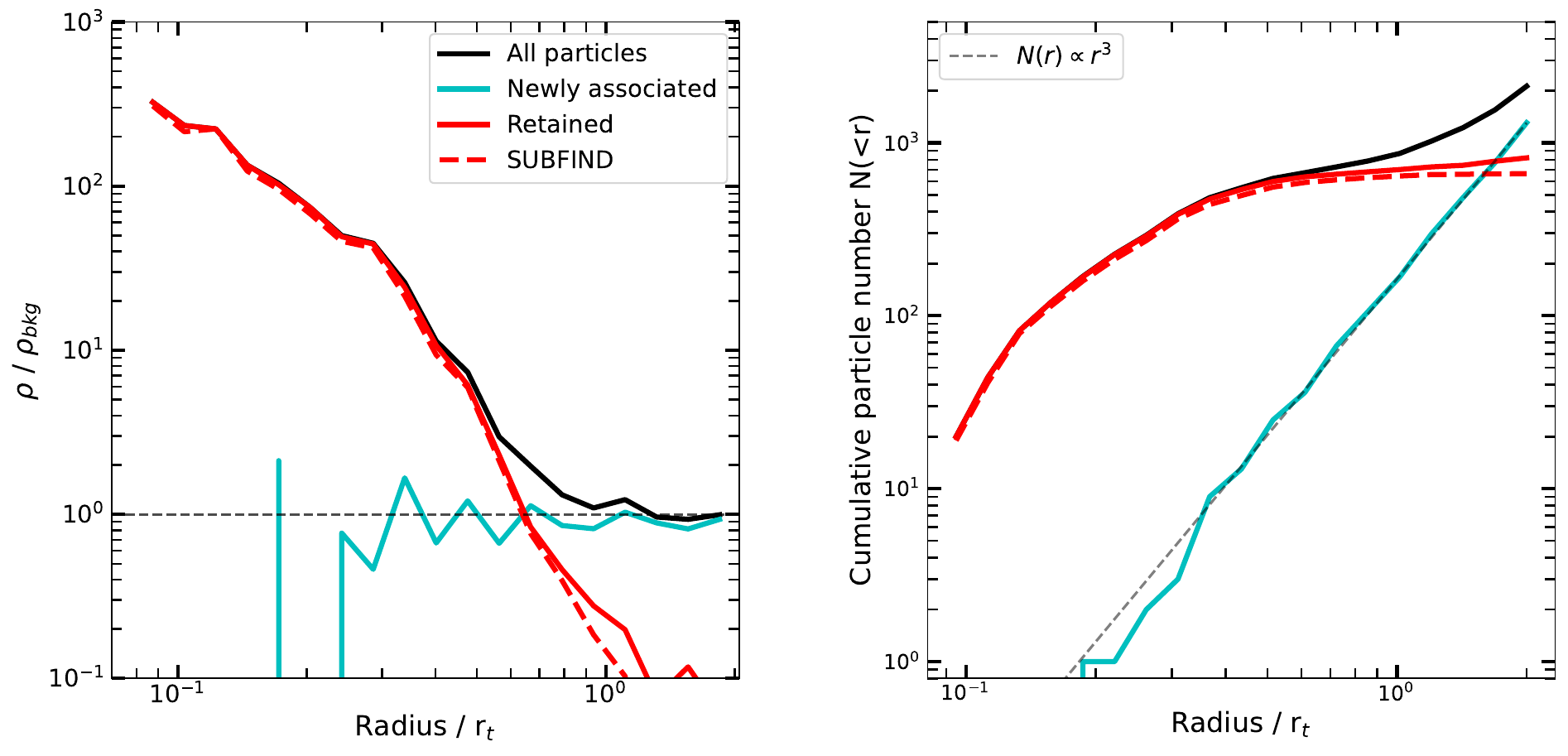}
	\centering
    \caption{Left panel: Density profiles for the subhalo of Fig.\ref{fig:sketch} at $z=0.11$ with radius normalised by $r_t$ and density normalised by $\rho_{\rm bkg}$, the mean density in the spherical shell $r_t<r<2r_t$. The black, red and cyan solid lines represent all, retained and newly associated particles, respectively. Additionally, a red dashed line shows the profile for particles identified by SUBFIND as bound to this subhalo. Right panel: The corresponding cumulative particle number profiles. The profile of newly associated particles increases as $r^3$ (the  black dashed line) indicating a nearly uniform density from the inner subhalo region out to the surrounding environment.}
    \label{fig:Randomcase}
\end{figure*}

If the subhalo were able to capture field particles temporarily, these might lead to an excess of newly associated particles  within the subhalo relative to the surrounding field. In Fig~\ref{fig:Randomcase} we test whether such temporary capture is significant by plotting the differential and cumulative density profiles of various particle populations at $z=0.11$. Black solid lines show  the profiles for all particles out to $2r_t$, while the red and cyan solid lines split these into retained and newly associated particles, respectively. The red dashed lines include only particles considered bound to the subhalo by SUBFIND; these are 81\% of the retained particles within $2r_t$ but almost 100\% of them below $0.5r_t$. Within $2r_t$ newly associated particles outnumber retained particles by a factor of about 1.5, but their relative contribution drops precipitously at smaller radii. In fact, their space density is everywhere consistent with that of the local field, as is seen most clearly through the close approximation of the cyan line in the right panel to the dashed black line indicating $N(<r)\propto r^3$.  These results are consistent with the idea that the subhalo just moves through the field without capturing particles or generating a significant local overdensity. If this were a starless subhalo, it would be very difficult to distinguish captured stars observationally from the background field population.

\begin{figure*}
	\includegraphics[width=0.8\textwidth]{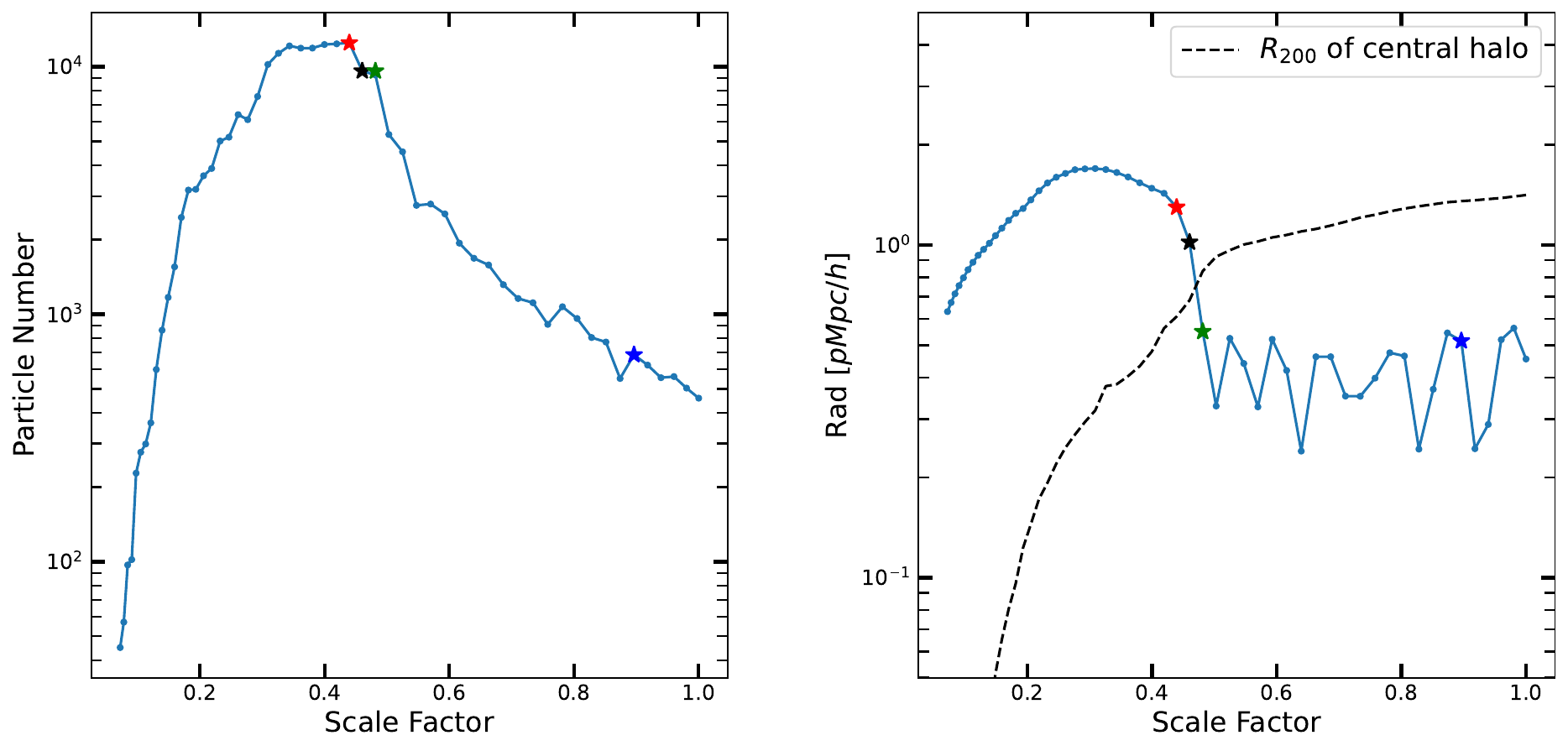}
	\centering
    \caption{Evolution of subhalo mass (left panel, indicated by SUBFIND bound particle number) and of distance to host halo centre (right panel) for the subhalo of Fig.\ref{fig:sketch}. In both panels, the black star marks the snapshot when the subhalo is last located outside the $R_{200}$ of host halo, which is also when set $A$ of associated particles is defined. The red star indicates $z_{\rm max}$, when $r_{\rm limit}$ is calculated, while the green star corresponds to $z_{\rm inf}$. We also marked 
    $z=0.11$ with a blue star.}
    \label{fig:orbit}
\end{figure*}

In Fig.~\ref{fig:orbit}, we  show the evolution of mass (characterized by SUBFIND bound particle number) and distance from host halo centre for this same subhalo. In both panels the black star indicates  $t_{\rm inf}$ while the red star indicates the time when set $A$ was defined. In the right panel, the black dashed line shows the evolution of $R_{200}$ for the host halo. The left panel of Fig.~\ref{fig:orbit} shows that the subhalo has lost 95\% of its infall mass by $z=0$. To characterize the orbit of the subhalo, we estimate the number of pericentric passages $N_{\rm orbit}$ by counting local minima of the blue curve in the right panel; $N_{\rm orbit}=6$ in this case.

\subsection{Stacked Density Profiles of Subhalos}

\begin{figure*}
	\includegraphics[width=\linewidth]{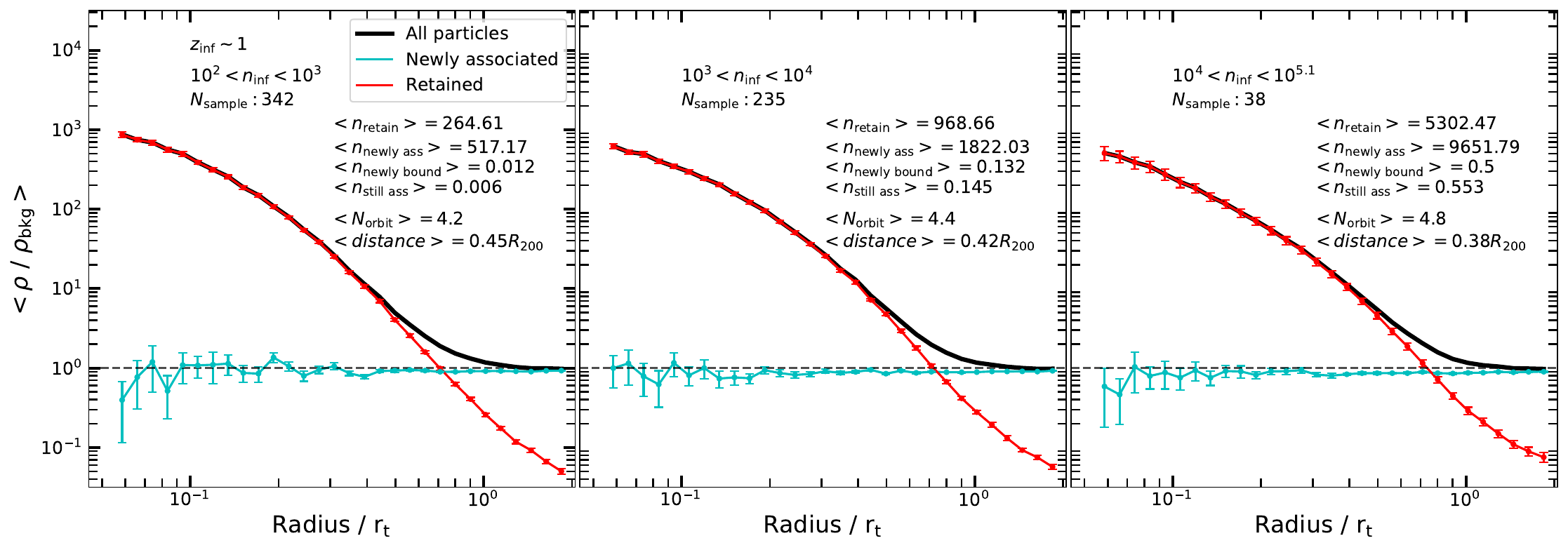}
	\centering
    \caption{Stacks of spherically averaged normalized density profiles at $z=0.11$ similar to those shown for an individual subhalo in Fig.\ref{fig:Randomcase}. The black, red and cyan solid lines represent straight averages for all, for retained and for newly associated particles, respectively. Error bars give the standard error in the mean estimated from the scatter among the profiles. From left to right, the panels show stacks for different ranges of $n_{\rm inf}$. Additional legends give the number of profiles in each stack. the mean numbers of retained and newly associated particles, the mean number of newly associated particles which are also bound (according to SUBIND), and the mean number of newly associated particles which are still associated at $z=0$. Two final legends give the mean number of orbits executed by $z=0.11$ and the mean halocentric distance of the subhaloes at this time.}
    \label{fig:averagez1}
\end{figure*}

\begin{figure*}
	\includegraphics[width=\linewidth]{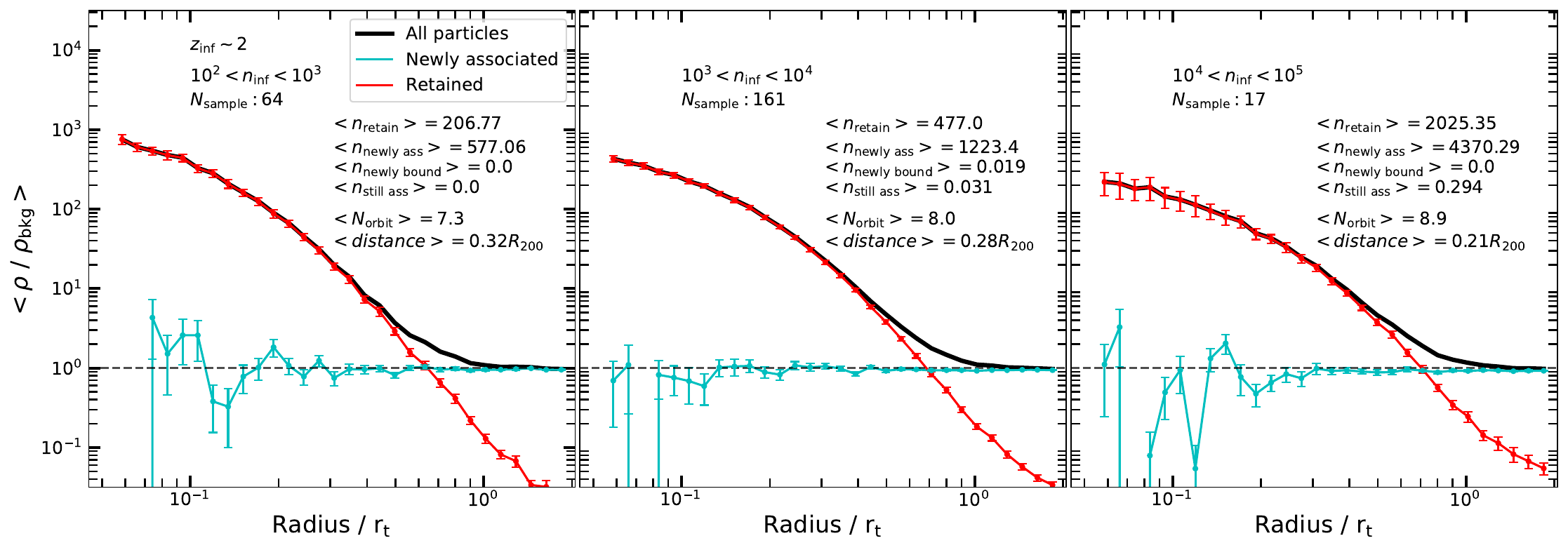}
	\centering
    \caption{As Fig~\ref{fig:averagez1}, but for $z_{\rm inf}\approx 2$}
    \label{fig:averagez2}
\end{figure*}

To obtain representative results based on all subhaloes in our Phoenix resimulations, we stack spherically averaged and normalised density profiles (exactly analogous to Fig.~\ref{fig:Randomcase}) as a function of infall mass and infall redshift for all $z=0.11$ subhaloes which still have 100 or more bound particles at $z=0$.  We divide these subhaloes into three mass bins according to their bound particle number at infall,: $10^{2}<n_{\rm inf}<10^{3}$, $10^{3}<n_{\rm inf}<10^{4}$ and $10^{4}<n_{\rm inf}<10^{5.1}$. Since most surviving subhalos were accreted onto their host after $z=2$ \citep[][]{2004MNRAS.355..819G,2015MNRAS.454.1697X} we here show results only for subhaloes with $z_{\rm inf} \approx 1$ and $z_{\rm inf} \approx 2$. For each case, three simulation snapshots close to the chosen redshift were considered.\footnote{Specifically, for $z_{\rm inf}\approx 1$ we used snapshots at $z = 1.08, 0.99$ and 0.91, while for $z_{\rm inf}\approx 2$ we used $z=2.07, 1.91$ and 1.77.} 

The mean density profiles of subhalos with $z_{\rm inf}\approx 1$ are shown in Fig.~\ref{fig:averagez1}. Black, red and cyan lines give results for all, for retained and for newly associated particles, respectively. The error bars on the red and cyan curves show the uncertainty in the mean derived from the scatter in the individual values in each bin. The number of  subhalo profiles averaged  is given as $N_{\rm sample}$ in each panel. In all cases the average result is very similar to that for the individual subhalo shown in Fig.~\ref{fig:Randomcase}; the density of newly associated particles is constant and equal to the surrounding ``field'' value all the way down to subhalo centre. There is no excess that might be attributed to temporarily or permanently captured particles.

A more direct way to look for newly captured particles is to see if any of the newly associated particles (i.e. the members of $B\setminus A$) is considered bound to the subhalo by SUBFIND. There are indeed a few such particles, but the average ($\langle n_{\rm newly~ bound}\rangle$ in the panels of Fig.~\ref{fig:averagez1}) is fewer than one per subhalo, about 0.01\% of the mean bound subhalo mass in all cases with no strong dependence on subhalo mass. Similarly one can check for captured particles independent of SUBFIND by finding the number of newly associated particles at $z\approx 0.1$ which are still associated at $z=0$ (i.e. particles which are members of $(B\setminus A)\cap C$). The mean number of such particles is given as $\langle n_{\rm still~assoc.}\rangle$ in the panels of Fig.\ref{fig:averagez1}, and is again less than one per subhalo, of order $10^{-4}$ of the subhalo mass.

We also give in Fig.~\ref{fig:averagez1}, the mean number  ($N_{\rm orbit}$) of orbits completed between $z_{\rm inf}$ and $z\approx 0.1$, and the mean distance of the subhaloes from halo centre at $z\approx0.1$. Surviving subhalos with lower $n_{\rm inf}$ (which have on average been reduced in mass by stripping by a smaller factor) have fewer orbits on average and are at larger distances from halo centre.

The stacked profiles for $z_{\rm inf}\approx 2$ are shown in Fig.~\ref{fig:averagez2}. These are very similar to those of Fig.~\ref{fig:averagez1} and again there is no detected excess of newly associated particles within $r_t$.
Subhalos with $z_{\rm inf}\approx 2$ execute more orbits by $z=0.1$, are closer to halo centre on average, and lose more mass on average than their counterparts with $z_{\rm inf} \approx 1$. This a natural consequence of the longer time they have spent within their host halo. Despite this longer residence time,
the values of $\langle n_{\rm newly~ bound}\rangle$ and $\langle n_{\rm still~assoc.}\rangle$ indicate that the fraction of the subhalo mass which has been captured after infall is still of order $10^{-4}$.

These results allow us to estimate  the mass in stars that might be captured by a starless subhalo in the Milky Way's halo. The stellar mass of the MW halo outside the Solar circle is somewhat less than $10^9M_\odot$ \citep{2019MNRAS.490.3426D}. Since the dark matter mass in the relevant region is a few times $10^{11}M_\odot$ the ratio of stellar mass to halo mass in the material captured by an orbiting starless halo will be of order 1000. We find the total captured mass to be about 0.01\% of a subhalo's current mass, so the current stellar to dark matter mass ratio for an initially starless subhalo can be estimated as $\sim 10^{-7}$. A starless subhalo of current mass $10^7M_\odot$ might thus have captured $\sim 1M_\odot$ of halo stars, a $10^8M_\odot$ subhalo might have captured $\sim 10M_\odot$. Galactic satellites with stellar masses almost this low have recently been identified, but it is unclear whether these are dark matter dominated \citep[e.g.][]{2024ApJ...961...92S,2025arXiv251011684T,2025arXiv251002431C}. All were found because they are very compact (half-light radii of a few parsecs) and very metal-poor. In the next section we study whether the captured population within an initially starless halo could plausibly be so compact and thus be identifiable against the uniform background of field stars that are just ``passing through''.  

\subsection{The distribution of captured particles}

\begin{figure}
	\includegraphics[width=\linewidth]{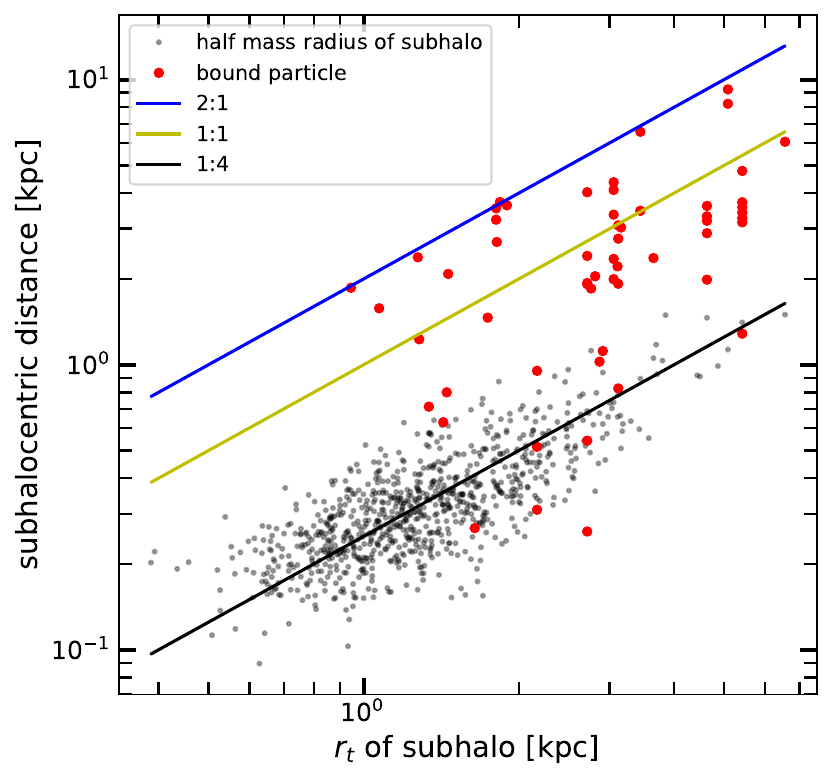}
	\centering
    \caption{Black points plot the half-mass radius of retained particles against tidal radius for each subhalo in the samples of Figures~\ref{fig:averagez1} and~\ref{fig:averagez2}. Red points plot the distance of newly captured particles from  subhalo centre against subhalo tidal radius. We reduce all lengths  by a factor of 10 to rescale our Phoenix haloes to the halo mass of Milky Way-like galaxies ($\sim 10^{12}M_\odot$).}
    \label{fig:radius}
\end{figure}


In Figure~\ref{fig:radius} red points show distance from subhalo centre as a function of subhalo tidal radius for all newly captured particles (i.e. those members of $B_i\setminus A_i$ that are bound to subhalo $i$ at $z=0.11$) for all the subhaloes in Figures~\ref{fig:averagez1} and~\ref{fig:averagez2}. Black points show the median distance of retained particles for these same subhaloes. This median distance is typically about $r_t/4$, as shown by the black line. The great majority of newly captured particles lie above this line with median distance closer to $r_t$ (the mustard line). Our definition of retained and newly associated particles requires them to lie within $2r_t$ (the blue line); since newly associated particles are almost uniformly distributed within this radius, their median distance is $2^{2/3}r_t$. Most newly captured particles are significantly closer to subhalo centre than this, but as a population they are substantially less centrally concentrated than retained particles.
Objects like Ursa Major III \citep{2024ApJ...961...92S} cannot be made of captured stars, since their characteristic property is that they are much {\it more} concentrated than the dark haloes they live in (if indeed they have one).

\begin{figure}
	\includegraphics[width=\linewidth]{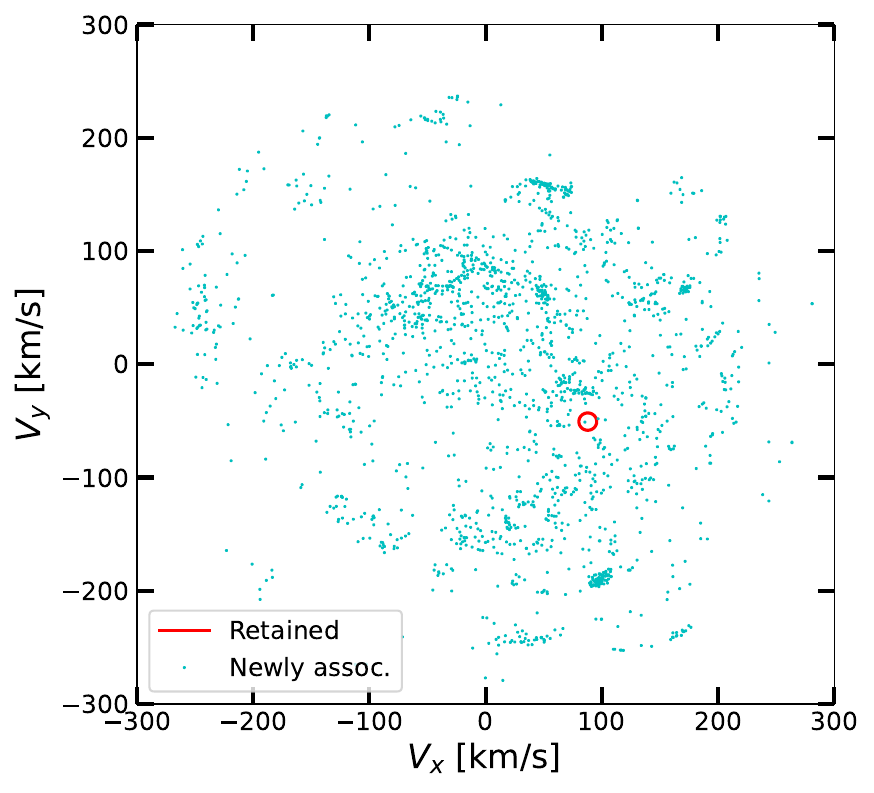}
	\centering
    \caption{Velocities relative to the main halo centre-of mass for newly associated particles at $z=0.11$  within the subhalo shown in Figures~\ref{fig:sketch},~\ref{fig:Randomcase} and~\ref{fig:orbit}. The velocities are scaled down by a factor of 10 to correspond to a halo of mass $\sim 10^{12}M_\odot$ similar to that of the Milky Way. The red circle is located at the velocity of the suhalo itself and has a radius which encloses all particles bound to it. Note the structure in the distribution of newly associated particles which includes several dense clumps corresponding to kinematically cold dark matter streams unassociated with this particular subhalo.}
    \label{fig:vel_res}
\end{figure}

The positions and velocities of originally starless haloes are, by definition, unknown, and these results show that they cannot be found using the positions of the $\sim 0.01\%$ of their newly associated stars which are actually captured. One might wonder if sufficiently accurate kinematic data would allow them to be found through clustering in velocity space.  Figure~\ref{fig:vel_res} shows that unfortunately this is not the case. Here we plot two components of the velocity in the rest-frame of the main halo for all particles at $z=0.11$ that are newly associated  with the particular subhalo of Figures~\ref{fig:sketch},~\ref{fig:Randomcase} and~\ref{fig:orbit}. The red circle indicates the subhalo velocity and has radius equal to its escape velocity. While the spatial distribution of newly associated stars is uniform and featureless within $2r_t$, the velocity distribution shows a remarkably rich structure including tight knots corresponding to kinematically cold streams of dark matter passing through the subhalo. No velocity-space feature is associated with the subhalo itself, as could have been anticipated from the fact that this particular subhalo had no truly captured particles. 

\section{Discussion} \label{sec:res}

The principal result of our analysis is that the capture of dark matter particles or stars by orbiting dark matter subhaloes is rare but does occur. For subhaloes which have completed at least three or four orbits within their host, the fraction of their bound mass acquired since
infall is about one part in 10,000, and does not appear to depend strongly on subhalo mass or on halo residence time. We have checked and find that roughly half of these captured particles were within twice the limiting radius we defined for identifying particles with the subhalo at infall (i.e. within $4R_{200}(z_{\rm max})$). Thus much of the captured material was plausibly part of a stream associated with the infalling halo, and so plausibly has few stars. Captured stars would have to have been formed at the centre of a different early halo which was subsequently tidally disrupted. Thus, the star to dark matter mass ratio of the 
captured material is likely to be smaller than that of the host halo as a whole.

As would intuitively be expected,  newly captured material is typically less strongly bound to a subhalo than material that has been retained since infall, and so is less strongly concentrated towards subhalo centre.  As a result, there is no detectable density contrast relative to the very much larger and effectively spatially uniform background of halo material that is just passing through the subhalo on unbound orbits. This material is almost all moving at relative velocities much larger than the escape velocity from the subhalo, but it has substantial structure in velocity space which precludes detection of the captured component unless the precise velocity of the subhalo is already known. Thus, initially starless subhaloes could not be detected through velocity space clumping of halo stars.

Out primary analysis has assumed that the results we have obtained for the Phoenix suite of cluster simulations can be applied (after scaling down by a factor of ten in size and velocity) to haloes of galaxies like the Milky Way. \cite{2012MNRAS.425.2169G} compare many properties of the Phoenix haloes to the 1,000 times less massive haloes of the Aquarius suite \citep{2008MNRAS.391.1685S}.
They find that the difference in mass manifests primarly in lower concentrations and formation times and in larger amounts of substructure in the Phoenix haloes. The shifts are relatively modest and should not significantly affect results for suhaloes of given (relative) mass and infall time. We have verified this (and also the lack of any significant resolution dependence in our primary results) by analysing the two highest resolution simulations of the Aquarius-A halo; details are given in Appendix A. 

Our results indicate that initially starless subhaloes will capture too few field stars 
to be distinguished against the much larger background of stars that are just passing through on almost unperturbed orbits. Because of the near-perfect scaling of these dynamical processes with halo mass, initially starless dark matter clumps in even smaller objects (for example, subhaloes in the haloes of dwarf satellite galaxies) will be even more difficult to detect, rarely capturing even a single star from the ambient population. Unfortunately, stellar capture is not a viable route for detecting initially starless dark matter clumps in any known system.

\section*{Acknowledgements}

We thank the referees for helpful comments which improved the paper. We are grateful to Volker Springel for reprocessing the Aquarius simulations and for helpful discussions. We thank the Aquarius and Phoenix collaborations for permission to use the simulations on which the results of this paper are based. HY and LG acknowledge support from the National Natural Science Foundation of China (Grant No. 12588202) and the National Key Research and Development Program of China (Grant No. 2023YFB3002500). In addition, HY thanks LACEGAL project with European Union’s HORIZON-MSCA-2021-SE-01 Research and Innovation program under the Marie Sklodowska-Curie grant agreement No. 101086388 for the grant which made possible the stay in Germany during which much of this work was carried out.

\section*{Data Availability}

The simulation data used in this article will be shared upon  reasonable request to the corresponding author.



\bibliographystyle{mnras}
\bibliography{example} 







\section*{Appendix A: tests of resolution and halo mass scaling using Aquarius haloes}
\label{Appendix A}

In this appendix we use two simulations from the Aquarius suite \citep{2008MNRAS.391.1685S} to test how our results depend on halo mass and on simulation resolution. Aq-A-2 is an isolated halo of mass, $M_{200c}=1.8\times 10^{12}M_\odot$. It was simulated with a similar zoom set-up and a similar number of particles to the Phoenix haloes analysed in the main text. This particular halo has a quiet history with an early formation time ($z_{\rm form}=1.9$, defined as the redshift when  $M_{200c}$ of the main progenitor is half the final value) and a high concentration ($c_{200c}= 16$) particularly when compared with the much more massive cluster haloes of the Phoenix suite (for which the median values of $z_{\rm form}$ and  $c_{200c}$ are 0.5 and 4.9, respectively). Aq-A-1 is a simulation of the same halo with 8 times higher mass resolution; \citet{2008MNRAS.391.1685S} show that most halo and subhalo properties are very well converged between these two simulations.

\begin{figure*}
	\includegraphics[width=0.58\linewidth]{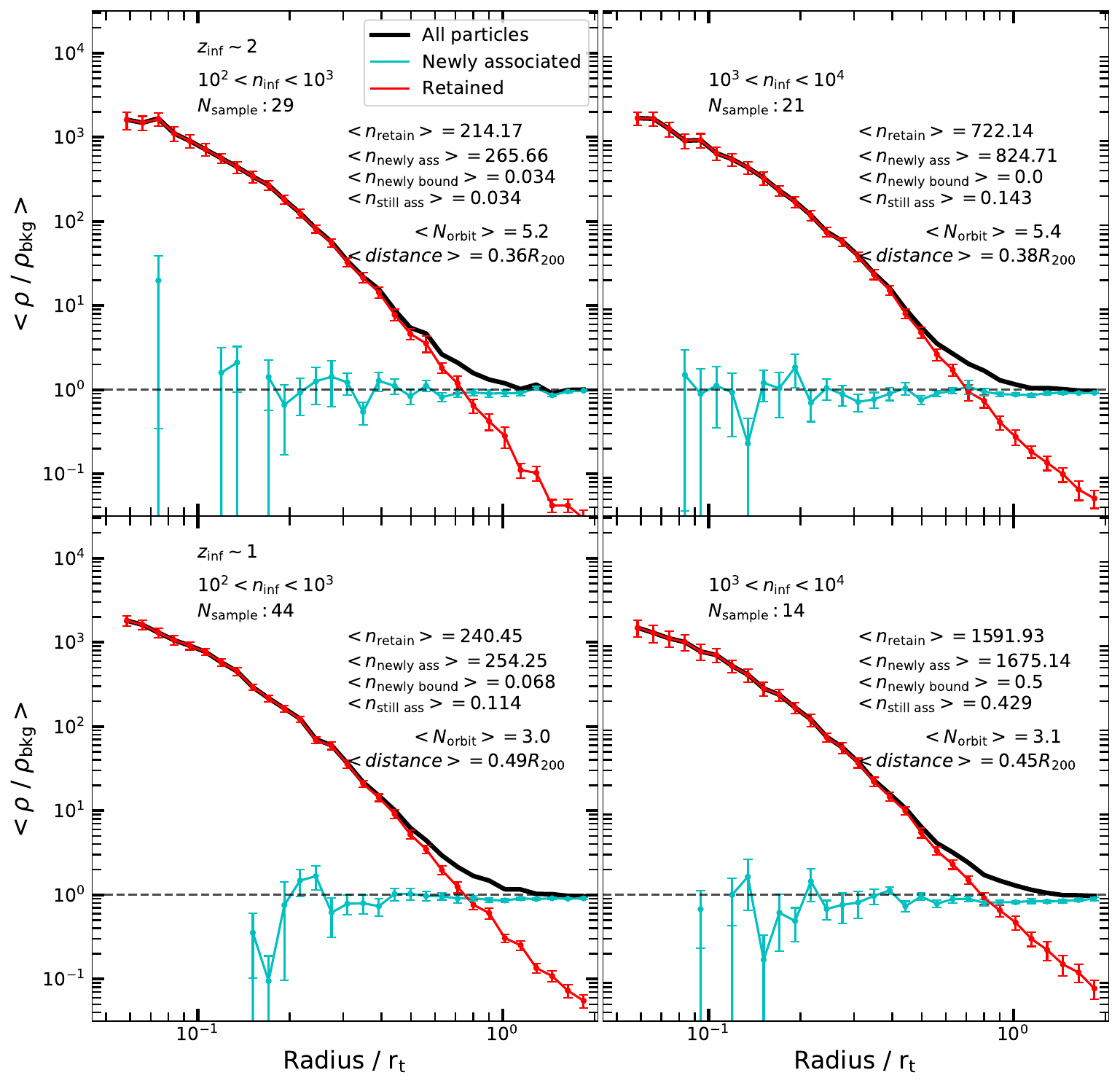}
	\centering
    \caption{Same as Fig~\ref{fig:averagez1}, with all definitions and legends unchanged, but for the Aq-A-2 simulation. From top to bottom, the panels show results at  $z_{\rm inf}\approx 1$ and $z_{\rm inf}\approx 2$, respectively.}
    \label{fig:Aq2}
\end{figure*}

\begin{figure*}
	\includegraphics[width=0.9\linewidth]{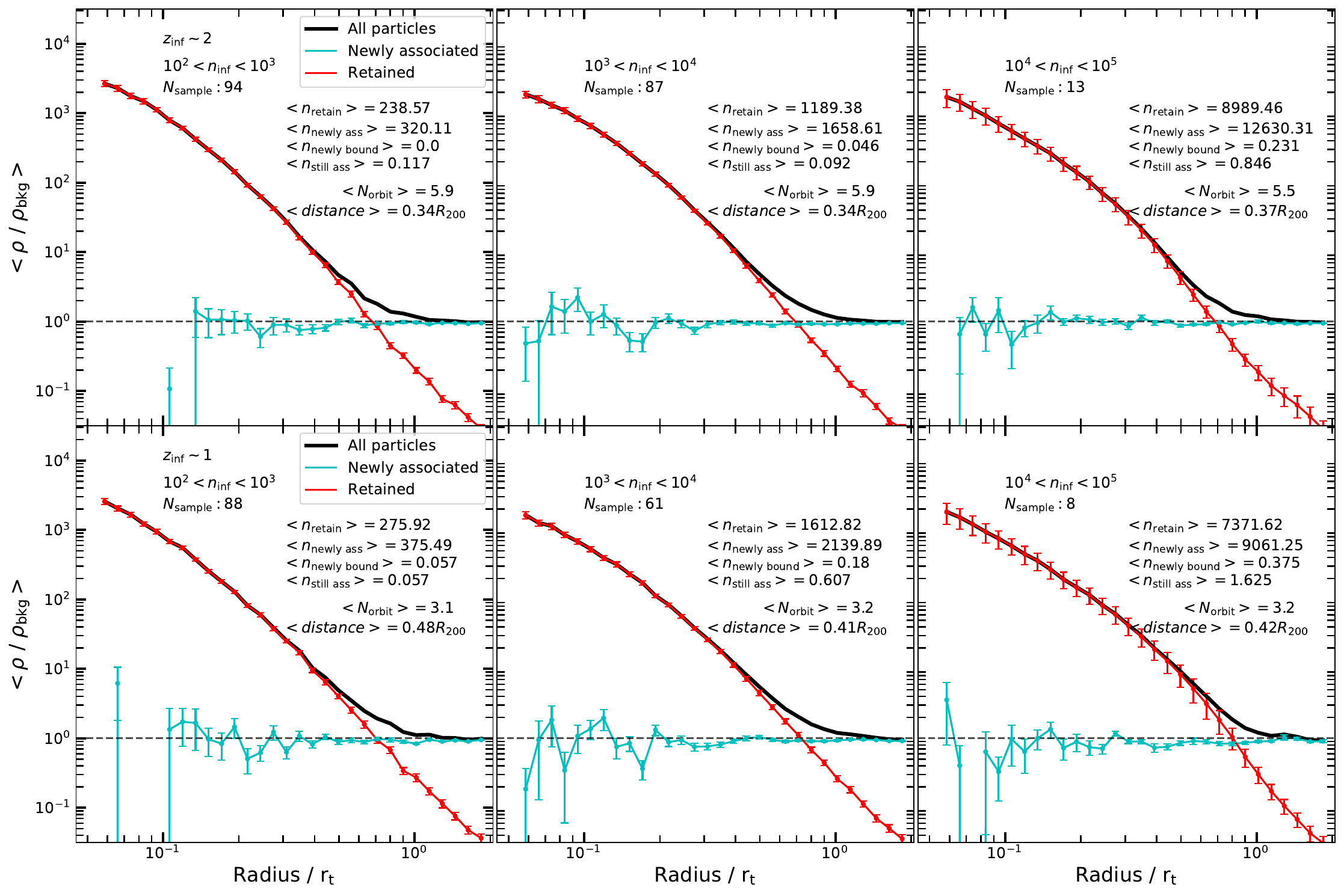}
	\centering
    \caption{Same as Fig~\ref{fig:Aq2}, but for the Aq-A-1 simulation.}
    \label{fig:Aq1}
\end{figure*}

We have repeated the analysis which led to Figures \ref{fig:averagez1} and \ref{fig:averagez2} for these two simulations and we present the results in figures \ref{fig:Aq2} and \ref{fig:Aq1}. Because of the higher concentration and earlier formation time of the Aquarius halo its subhalo population is shifted to substantially lower (relative) masses than in the Phoenix simulations. As a result,
only the two lower ranges of $n_{\rm inf}$ are populated in Aq-A-2. The higher mass resolution of Aq-A-1 results in all 3 ranges being populated, but the subhalo infall massses corresponding to the range  limits are now 8 times lower. 

Both qualitatively and quantitatively, the results for these Aquarius haloes are very similar to those shown in the main text for the Phoenix haloes. Despite the large differences in halo mass, formation time and concentration, newly associated particles are uniformly distributed across each subhalo and are almost all unbound. The fraction of subhalo mass that is truly captured is unchanged, about $10^{-4}$ both for $z\sim 2$ and for $z\sim 1$, again with no obvious dependence on subhalo mass. The only clear systematic difference with the Phoenix case is that surviving subhaloes in Aquarius-A have, on average, completed fewer orbits since infall than in Phoenix. Results are well converged between Aq-A-2 and Aq-A-1 showing that particle discreteness has no substantial effect on our results.
Thus it appears that the principal results of this paper are almost independent of halo mass and concentration and so can be safely extrapolated over the full range of halo masses.

\section*{Appendix B: A toy model for the stellar halo of the Milky Way}

Equation (\ref{eq:MLarg}) shows that, in the mean,  the stellar mass accreted onto a initially starless subhalo as it orbits through a halo containing both dark matter and stars is equal to the accreted dark matter mass multiplied by $\langle m_*/m_{\rm dm}\rangle$, where the average is taken over all orbits from which accretion can occur.  In galaxy haloes a substantial fraction both of the mass and of the stars is in the form of streams, so that mass and stars are not necessarily well-mixed along any particular orbit.\footnote{Note that our derivation of equation (\ref{eq:MLarg}) is completely general and does not require equilibium nor that stars and dark matter be identically and smoothly distributed in phase along an orbit.}. However, since the subhalo under consideration is assumed to be initially starless, all accreted stars must come from streams other than that of the subhalo itself, and so the orbital phase from which capture can occur is distributed along the orbit in a way which is at most weakly correlated with the phase distributions of stars and dark matter.\footnote{Note further, that capture is strongly dependent both on the integrals of motion and on the orbital phase of the accreted material While the former are strongly correlated for a given stream, the latter spread widely. Thus capture probabilities will be at most weakly correlated in different parts of a stream.} Thus, the star-to-dark matter ratio of captured material should still be $\langle m_*/m_{\rm dm}\rangle$ in the mean. In this appendix we present a simple but moderately realistic toy model in which it is easy to calculate the relation between the accretion-orbit-averaged $\langle m_*/m_{\rm dm}\rangle$ of Equation (\ref{eq:MLarg}) and the directly observable local ratio in the halo at the mean radial position of the subhalo.

The rotation curves of spiral galaxies are approximately flat at the halo radii of interest, so a simple but appropriate dark matter halo model is a singular isothermal sphere with circular velocity $V_c$, for which the enclosed mass, the density and the gravitational potential are given by
\begin{equation}
    M(r) = rV_c^2/G,~~ \rho(r) =\rho_0(r/r_0)^{-2}~\textrm{and}~\Phi(r)=V_c^2\ln(r/r_0),
\end{equation}
 where $r_0$ is a reference radius and $\rho_0=V_c^2/(4\pi G r_0^2)$. The (isotropic) phase-space distribution function for the dark matter is a function of $E\equiv \Phi(r)+v^2/2$ alone and is simply
 \begin{equation}
 f_{\rm dm}(E) = \rho_0 (\pi V_c^2)^{-3/2} \exp(-2E/V_c^2).
 \end{equation}
 
 The luminosity profiles of stellar halos are typically steeper than their dark matter density profiles. For example, around the Milky Way the density profile of the stellar halo is well represented by a power law $\rho_*\propto r^{-\gamma}$ with  $\gamma\approx 3$ out to about 30~kpc \citep{2018MNRAS.474.2142I}, hence in the region where a poorly populated low-mass dark matter halo might be observable. For an isotropic population within our isothermal halo, this implies a stellar distribution function
 \begin{equation}
 f_*(E) = \rho_{*,0} (2\pi V_c^2/\gamma)^{-3/2} \exp(-\gamma E/V_c^2),
 \end{equation}
 where $\rho_{*,0}$ is the local mass density of the stellar halo  at $r_0$. The mean stellar-to-dark matter mass ratio on orbits of energy $E$ is thus
 \begin{equation}
     \langle m_*/m_{\rm dm}\rangle(E) = (\rho_{*,0}/\rho_0) (\gamma/2)^{-3/2} \exp[(2-\gamma)E/V_c^2], 
     \label{eq:MLarg1}
 \end{equation}
 where $m_*/m_{\rm dm} = (\rho_{*,0}/\rho_0)(r/r_0)^{2-\gamma}$ then gives the local star-to-dark matter mass ratio as a function of radius.

 At the time a halo star is captured by a low-mass orbiting subhalo the positions of the two must coincide within a distance comparable to the subhalo tidal radius $r_t$ and their velocities must differ by no more than the subhalo escape velocity $v_e$. Since $r_t\ll r$ and $v_e\ll V_c$, the star and the subhalo must have very similar $E$ at the moment of capture. Thus the capture-orbit-averaged $\langle m_*/m_{\rm dm}\rangle$ of equation (\ref{eq:MLarg}) can be taken to be $\langle m_*/m_{\rm dm}\rangle(E_{\rm sub})$ from equation~(\ref{eq:MLarg1}) where $E_{\rm sub}$ is the orbital energy of the subhalo. Possible positions for a subhalo of this energy oscillate between the corresponding apo- and pericentres and bracket the circular orbit radius $r_{\rm circ}(E_{\rm sub})$, given by $E_{\rm sub} = V_c^2(\ln (r_{\rm circ}/r_0) + 1/2)$. Substituting this expression into equation (\ref{eq:MLarg1}) shows that the mean $m_*/m_{\rm dm}$ of captured material differs from the local value at $r_{\rm circ}$ by a factor of $(\gamma/2)^{-3/2}\exp(1-\gamma/2))$ which evaluates to 0.33 for $\gamma=3$. Thus in a system analogous to the Milky Way's halo, the star-to-dark matter mass ratio of the material accreted by a subhalo is expected to be a few times smaller than the local value at a typical position along the subhalo's orbit. This suggests that the arguments in our main text give rise to a mild overestimate of the mean number of stars which will be captured by starless subhaloes.

 Although the simulation analysis of our main paper and of Appendix A  gives a robust estimate of the {\it mean} mass accreted by a {\it population} of subhaloes, the real stellar and dark matter distributions are both (to differing degrees) made up of a superposition of streams, rather than having smooth, well-mixed distribution functions like those assumed in this appendix. As a result, the scatter in the dark matter and stellar masses accreted by individual subhaloes may be substantial. Very occasionally, one particular subhalo may accrete significantly more than its ``fair share" of stars because its orbit happens to align with a rich stellar stream\footnote{A particularly extreme version of this scenario occurs when a low-mass  starless subhalo merges with a more massive subhalo which does contain stars. As noted in the main text, this possibility was discussed by \citet{2023ApJ...958..166W} and we excluded a few cases of such subhalo mergers from our own analysis.}; in compensation, a much larger number of similar subhaloes must then accrete fewer stars than expected so that the overall population mean matches the small stellar mass we have estimated from our simulations. The size of such effects can only be evaluated through realistic simulations which resolve the fine-grained distributions of both stars and dark matter. This is well beyond the resolution even of the Aq-A-1 simulation analysed in Appendix A, which remains the highest resolution, dark-matter-only simulation of a Milky Way-like galaxy halo published so far.

\label{Appendix B}

\bsp	
\label{lastpage}
\end{document}